\documentclass[aps,prl,twocolumn,showpacs]{revtex4}
\usepackage{graphicx}
\usepackage{amsmath}
\usepackage{amssymb}
\usepackage{bm}
\usepackage{dcolumn}
\usepackage{color}

\begin{document}

\title{Quantum Noise-to-Sensibility Ratio}

\author{B. M. Escher}
\email{bmescher@if.ufrj.br}
\affiliation{Instituto de F\'{\i}sica, Universidade Federal do Rio de Janeiro, 21.941-972, Rio de Janeiro (RJ) Brazil}

\date{\today}

\begin{abstract}
The quantum variables that can be accessed directly by experiments are described by observables. Therefore, physical parameters can only be evaluated indirectly, via estimations based on experimental measurement results. I show that the quantum sensitivity, or the quantum statistical uncertainty  in single-parameter estimation, can be defined as the (minimal) ratio between noise and sensibility with a parameter of a well-calibrated observable. Among one of its applications, I show that, measuring a convenient quadrature in squeezed probe oscillators, it is possible to surpass the standard precision limit in phase-shift estimations, even in presence of moderate phase diffusion.
\end{abstract}


\maketitle

\textit{Introduction}.---The estimation of parameters in the framework of quantum physics is a very challenging problem \cite{Helstrombook76,Braunsprl94,braum96}. Quantum features of probe systems, for instance, may allow enhancements to statistical uncertainty of estimations, making possible to arrive at a determined level of precision with less amount of physical resources \cite{Giosci04,natexp}. For reasonable (unbiased) estimators \cite{Cramerbook46}, a way to foresee these enhancements theoretically is via the quantum Fisher information (QFI) \cite{Braunsprl94,braum96}, since the inverse of the square root of QFI delimits the sensitivity. This limit, founded with QFI, bounds the performance of any strategy of estimation, including best estimators and more informative experimental measurement devices. For ideal situations, QFI can be calculated analytically as a function of pure probe states. However, for more realistic scenarios, where probes are interacting with non-monitoring environments, the analysis of possible enhancements provided by quantum strategies is more involved \cite{Gionpho11}. Up to now, there is no practical expression for QFI as a function of mixture probe states in a given pure state decomposition \cite{com1}. In trying to circumvent this problem, it is shown in Refs. \cite{Eschernp11,Escherxiv12} a variational approach to calculate the QFI via upper bounds. Therefore, it is possible, with this approach, to delimit the quantum enhancement of the precision in a parameter estimation under the presence of external noise. On the other hand, a practical answer to the question of how to get the QFI via lower bounds remains undeveloped so far. Besides to complement that variational approach, this point is important because it may guarantee some quantum enhancement in the sensitivity even in non-ideal and non-optimal situations.

An inherent property of quantum theory is that all dynamical variables, which can be measured directly, are described by self-adjoint operators called observables. For this reason, parameters characterizing physical processes cannot be accessed straight by experiments, they can only be estimated based on the experimental data. A practical way to describe the sensitivity with a parameter in this type of problem is considering the uncertainty formula obtained for indirect measurements, given by the combined standard uncertainty \cite{jcgm}. In quantum physics, it is the ratio between noise and sensibility with the parameter of the measured observable. Then, the best strategy in this picture is measuring an optimum observable that minimizes this ratio. An unanswered question in this analysis, as far as I know, is about the relation between this sensitivity, based on the minimal noise-to-sensibility ratio, and the one obtained with QFI.

In this Letter, I present a solution to those two unaddressed questions, showing that QFI can be defined also as the maximum of the square of the inverse of the noise-to-sensibility ratio over all physical observables. Moreover, the symmetric logarithmic derivative operator (SLD) \cite{Helstrombook76, Braunsprl94,braum96} minimizes this ratio, being therefore an optimum observable to be measured. These results imply that, at least in the regime of large samples, the unknown physical parameter can be estimated efficiently, considering, in average, only the observed values of a well calibrated SLD. This solution also constitutes a prescription for evaluating QFI via lower bounds, which is relevant for practical purposes, when analytical calculus are required.

To illustrate the potential of this approach, I also tackle here the problem of estimation of a phase-shift with quantum-probe oscillators under phase diffusion. This question has been addressed recently both theoretically \cite{Escherxiv12,Parisprl11} and experimentally \cite{Parisexp}. In Ref.~\cite{Parisprl11}, the sensitivity is calculated numerically for initial Gaussian states of the probe and, from this, the ultimate precision limit is obtained for these states as a function of the average energy of the probe. In Ref.~\cite{Parisexp}, an experiment is done with coherent states, reaching approximately to the corresponding standard limit to the sensitivity. Finally, in Ref.~\cite{Escherxiv12}, lower bounds for the sensitivity are derived analytically, and are valid for any initial state of the probe, showing that the sensitivity cannot be smaller than a noise-diffusion constant. I show here that, even in the regime of moderate phase diffusion, quadrature measurements is almost optimal for phase-shift estimation with Gaussian states and, with this measurement, it may be possible to surpass the standard sensitivity limit by squeezing the initial state of the probe oscillator. 

\textit{Quantum noise-to-sensibility ratio}.---In the problem of estimation of a parameter with quantum probes, it is generally assumed that (i) the unknown value of the parameter, which is aimed to be estimated, is $x_{\rm true}$; (ii) this value is within a known interval of real numbers $x$; and (iii) the $x$-dependent physical process changes the state of the quantum probe to $\hat\rho(x)$. Notice that the true state of the probe, changed by the physical process, is $\hat\rho(x_{\rm true})$, which is unknown because $x_{\rm true}$ is unknown.

\begin{figure}[t]
\centering
\includegraphics[width=0.45\textwidth]{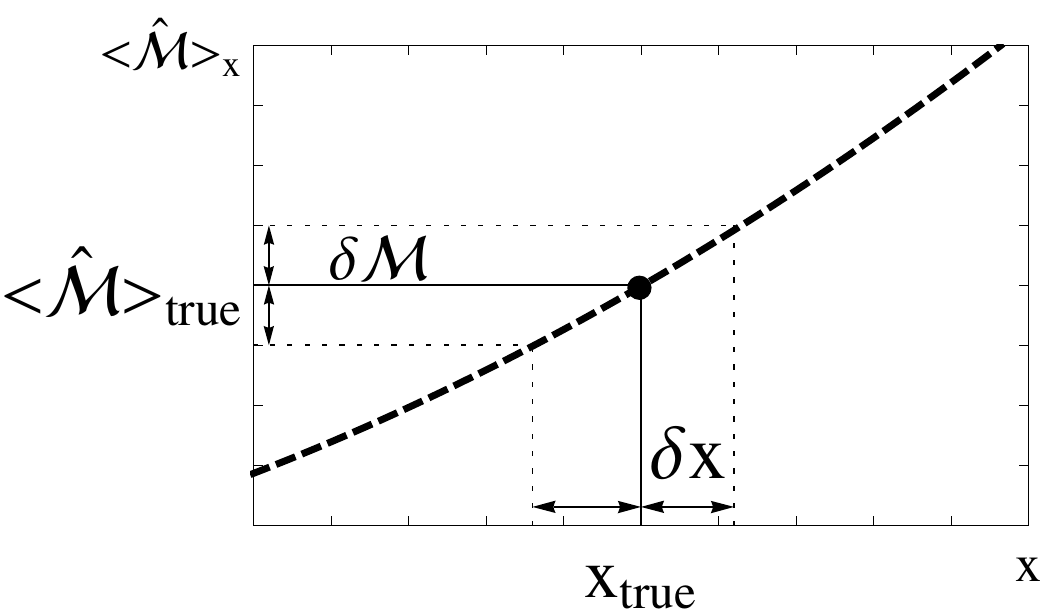}
\caption{The average value of the observable $\hat{\cal M}$,$\langle\hat{\cal M}\rangle_{x}$, versus the possible values $x$ of the parameter. This function is shown by the dashed line. The relation between the uncertainty $\delta{\cal M}$ and $\delta x$ is illustrated by the dotted lines, while the full lines link $\langle\hat{\cal M}\rangle_{\rm true}$ to $x_{\rm true}$. By analogy, the estimator that connects any value $\langle\hat{\cal M}\rangle_{x}$ to $x$ can be defined in the same manner \cite{estc}.}
\label{fig1}
\end{figure}

The information about the value of the parameter is hidden in the properties of the probed state $\hat\rho(x_{\rm true})$. A way to get this information is measuring some observable $\hat{\cal M}$ of the probe, which allows one to differentiate the possible states $\hat\rho(x)$ and to estimate the value of the parameter based on the observed measurement results. In principle, this strategy will always work when the average value of the observable $\hat{\cal M}$ is a bijective function of $x$. This function is denoted here, for convenience, as $\langle\hat{\cal M}\rangle_{x}$, where $\langle\bullet\rangle_{x}:={\rm Tr}[\hat\rho(x)\bullet]$. That works because, if the true average $\langle\hat{\cal M}\rangle_{\rm true}$ is measured, $x_{\rm true}$ can be estimated exactly by inverting the known function $\langle\hat{\cal M}\rangle_{x}$ in the observed point, as shown in Fig.~\ref{fig1}. However, in very general situations, there is a fluctuation, an error $\delta\cal{M}$ in the experimental observed value $\langle\hat{\cal M}\rangle_{\rm exp}$, i.e. $\langle\hat{\cal M}\rangle_{\rm exp}=\langle\hat{\cal M}\rangle_{\rm true}\pm\delta\cal{M}$, which yields, by propagation, an uncertainty in the estimation of the parameter. There are two main sources of noise for this error, one from the intrinsic probabilistic nature of quantum physics, since the state $\hat\rho(x_{\rm true})$ may have no definite value for the observable $\hat{\cal{M}}$, and another from spurious fluctuations, provoked by the interaction between the probe and the environment, which is not monitored. Usually, this second source of noise has also a statistical, a random nature, enabling a perfect estimation only in the asymptotic regime of infinite samples. 

In the formalism of quantum physics, the uncertainty $\delta\cal{M}$ is described by the standard deviation of $\hat{\cal M}$,
\begin{equation}\label{sdev}
\delta{\cal M}:=\sqrt{\langle{\cal M}^2\rangle_{x_{\rm true}}-\langle\hat{\cal M}\rangle^{2}_{x_{\rm true}}}\equiv\sqrt{\langle\Delta\hat{\cal M}^2\rangle_{x_{\rm true}}}.
\end{equation} 
As discussed above, this uncertainty yields a finite sensitivity with $x$. The dependence of this sensitivity with the chose observable $\hat{{\cal M}}$ appears explicit in the formula of combined standard uncertainty \cite{jcgm}. As justified in Fig.~\ref{fig1}, it is given by the noise-to-sensibility ratio:
\begin{equation}\label{nsr}
\delta x_{\rm nsr}:=\left.\dfrac{\sqrt{\langle\Delta\hat{{\cal M}}^2\rangle_{x}}}{\left\vert d\langle\hat{{\cal M}}\rangle_{x}/dx\right\vert}\right\vert_{x\to x_{\rm true}},
\end{equation}
which allows an accurate estimation of the value of the parameter if the experimental data of $\hat{{\cal M}}$ has small uncertainty and high sensibility with the parameter.

Equation~\eqref{nsr} is demonstrated rigorously with the condition that $\delta{\cal M}$ is a ``small" quantity. Based on the expansion of the estimator defined according to Fig.~\ref{fig1}, see also Ref.~\cite{estc}, up to second order in $\delta{\cal M}$, this  condition reads $\delta{\cal M}\ll2(d\langle\hat{\cal M}\rangle_{x_{\rm true}}/dx_{\rm true})^2/\vert d^2\langle\hat{\cal M}\rangle_{x_{\rm true}}/dx^2_{\rm true}\vert$. Beyond that, it may be used as a necessary condition to attain the regime of large samples $\nu$, since $\delta{\cal M}\sim1/\sqrt{\nu}$. Notice that the present estimator is unbiased in this limit. Notwithstanding its regime of application, equation~\eqref{nsr} can be used in general as a first figure of merit for characterizing the sensitivity in the estimation of a parameter based on the average values of $\hat{\cal M}$. 

The noise-to-sensibility ratio of an observable $\hat{\cal M}$ in the estimation of a parameter with quantum probes is given by Eq.~\eqref{nsr}. The square of its inverse provides a measure of how much information it is possible to gain about the parameter with this strategy of inferring the value of the parameter by measuring $\hat{\cal M}$. Therefore, the quantum noise-to-sensibility ratio is defined as the minimum of \eqref{nsr} over all observables, and the correspondent maximal amount of information ${\cal F}_{\rm Q}^{\rm nsr}$ is defined similarly as
\begin{equation}\label{fsnr}
{\cal F}_{\rm Q}^{\rm nsr}(x_{\rm true}):=\max_{\hat{\cal M}}\left\{\left.\dfrac{\left[ d\langle\hat{{\cal M}}\rangle_{x}/dx\right]^2}{\langle\Delta\hat{{\cal M}}^2\rangle_{x}}\right\vert_{x\to x_{\rm true}}\right\}.
\end{equation} 
Notice that the optimum observable that maximizes the right hand side of Eq.~\eqref{fsnr} may depends on $x_{\rm true}$, but not on $x$. To accomplish a concrete experiment, it is necessary to choose an observable and, once the parameter is unknown, this observable would be, in principle, the same for different values of the parameter. Of course that this reasoning does not forbid one to consider adaptive strategies, in which the procedure will depend explicitly on the possible measured values, and not on $x$.

The value of ${\cal F}_{\rm Q}^{\rm nsr}(x_{\rm true})$ and the optimum operator $\hat{\cal M}_{\rm opt}$ that maximizes the right hand side of Eq.~\eqref{fsnr} can be found via variational methods. Taking $\hat{\cal M}=\hat{\cal M}_{\rm opt}+\delta\hat{\cal M}$ and expanding the right hand side of \eqref{fsnr} in first order of the Hermitian operator $\delta\hat{\cal M}$, it is possible to determine an equation for $\hat{\cal M}_{\rm opt}$ such that the corresponding variation is zero for any $\delta\hat{\cal M}$. It is obtained then straightforwardly:
\begin{widetext}
\begin{equation}\label{mopt}
\dfrac{\hat\rho(x_{\rm true})[\hat{\cal M}_{\rm opt}-\langle\hat{\cal M}_{\rm opt}\rangle_{x_{\rm true}}]+[\hat{\cal M}_{\rm opt}-\langle\hat{\cal M}_{\rm opt}\rangle_{x_{\rm true}}]\hat\rho(x_{\rm true})}{2}=\dfrac{\langle\Delta\hat{\cal M}_{\rm opt}^{2}\rangle_{x_{\rm true}}}{{\rm Tr}[(d\hat\rho(x_{\rm true})/dx_{\rm true})\hat{\cal M}_{\rm opt}]}\dfrac{d\hat\rho(x_{\rm true})}{dx_{\rm true}}.
\end{equation}
\end{widetext}

This equation admits more than one solution for $\hat{\cal M}_{\rm opt}$. It is possible to understand this non-uniqueness by analyzing more closely Eq.~\eqref{nsr}. That equation is invariant via the following transformation: $\hat{\cal M}\to a(\hat{\cal M}-b)$, where the real numbers $a$ and $b$ are independent on $x$. As expected, the noise-to-sensibility ratio does not depend on either of the units, or the dimension of the observable in question (expressed by $a$) nor on a fixed bias of $\hat{\cal M}$ (expressed by $b$). With this freedom, one can set up $\langle\hat{\cal M}_{\rm opt}\rangle_{x_{\rm true}}=0$, and work with natural units of~$1[x^{-1}]$, in which case is valid the relation $\langle\Delta\hat{\cal M}_{\rm opt}^{2}\rangle_{x_{\rm true}}={\rm Tr}[(d\hat\rho(x_{\rm true})/dx_{\rm true})\hat{\cal M}_{\rm opt}]$. For this choice, Eq.~\eqref{mopt} is simplified to a Sylvester-like equation \cite{bartels}, being the same one satisfied for the SLD, wrote as $\hat{L}(x_{\rm true})$. Therefore, this operator is an optimum observable to be measured. Moreover, the amount of information obtained with this procedure, calculated by taking $\hat{L}(x_{\rm true})$ into Eq.~\eqref{fsnr}, is the maximal one allowed by quantum physics for unbiased estimators, given by QFI:
\begin{equation}\label{fq}
{\cal F}_{\rm Q}^{\rm nsr}(x_{\rm true})=\langle\hat{L}^{2}(x_{\rm true})\rangle_{x_{\rm true}}.
\end{equation}

This is the main result of this Letter: showing that, at least in the regime of large samples, in which Eq.~\eqref{nsr} is attainable, the quantum sensitivity can be obtained by measuring the SLD, if calibrated correctly, since $x_{\rm true}$ is unknown, and the estimative can be based solely on the experimental average value of this observable. Therefore, the problem of finding the measurement and the estimator that yields to the quantum sensitivity regime is reduced, in principle, to a calibration problem. These results allow also interpreting physically the optimum SLD: its eigenvectors determine an optimum base of measurements and its eigenvalues have already the best statistical treatments for the possible experimental data. Furthermore, equality \eqref{fq}, together with definition \eqref{fsnr}, yields a practical approach for determining QFI via lower bounds. Whenever Eq.~\eqref{mopt} is too hard to be solved, it can be used to give an ansatz for $\hat{\cal M}_{\rm opt}$. So that even if the guessed observable is not the optimal one, it can still be used to calculate the square of the inverse of the noise-to-sensibility ratio, yielding a lower bound to QFI.

The optimal observable to be measured depends usually on the value of the unknown parameter \cite{braum96, gill}. This problem can be solved with (i) a proper experimental device, which can be tuned, calibrated in such a way that it performs a measure of $\hat{L}(x_{\rm exp})$ for any $x_{\rm exp}$ within the interval $x$; and (ii) an adaptive strategy, which changes conveniently the values of $x_{\rm exp}$ according to the observed results. The number of necessary trials for calibrating this experimental device within this approach can be estimated by expanding the function ${\cal F}^{\rm nsr}(x_{\rm exp})$, which is the square of the inverse of the noise-to-sensibility ratio of $\hat{L}(x_{\rm exp})$, around $x_{\rm true}$. Indeed, taking this expansion up to second order in $x_{\rm true}-x_{\rm exp}$,
\begin{equation}\label{fg}
{\cal F}^{\rm nsr}(x_{\rm exp})\!=\!{\cal F}_{\rm Q}^{\rm nsr}(x_{\rm true})\!-\!{\cal G}(x_{\rm true})\delta x_{\rm exp}^{2}\!+\! {\cal O}(\delta x_{\rm exp}^{3}),
\end{equation}
where $\delta x_{\rm exp}:=x_{\rm true}-x_{\rm exp}$ and
\begin{equation}
{\cal G}(x)\!\!:=\!\!\langle[\Delta(d\hat{L}(x)\!/\!dx)]^2\rangle_{x}\!\!-\!\langle d\hat{L}^2(x)\!/\!dx\rangle_{x}^2/[4\langle\hat{L}^2(x)\rangle_{x}],
\end{equation}
and taking also the inequality $\delta x_{\rm exp}^{2}\ge1/[\nu{\cal F}_{Q}^{\rm snr}(x_{\rm true})]$ \cite{inec}, this condition reads 
\begin{equation}\label{nucon}
\nu\gg{\cal G}(x_{\rm true})/[{\cal F}_{Q}^{\rm nsr}(x_{\rm true})]^2.
\end{equation}

When the parameter to be estimated comes from a unitary process $\hat{U}(x)=\exp{(-ix\hat{h})}$, where $\hat{h}$ is a Hermitian operator that does not depend on $x$, and the state of the probe is pure, such that $\hat{\rho}(x)=\hat{U}(x)\vert\psi\rangle\langle\psi\vert\hat{U}^{\dagger}(x)$, QFI can be calculated analytically, being equal to $4\langle\Delta\hat{h}^2\rangle$. While \eqref{nucon} simplifies to $4\nu\!\!\gg\!\![\langle(\Delta\hat{h})^4\rangle/\langle\Delta\hat{h}^2\rangle^{2}-1]$. Interestingly, for some classes of states $\vert\psi\rangle$, the same condition was founded in Ref.~\cite{bra94}, except for a multiplicative factor of $2$. There, this is done from the minimum sample size that makes the maximum likelihood estimator to be approximately Gaussian \cite{bra92}.  Here, condition \eqref{nucon} is based on the number of trials that are necessary for calibrating an optimum measurement device.

\textit{Phase-shift estimation under phase diffusion}.---The goal is to estimate a phase-shift $\phi_{\rm true}$ in a state space of a quantum oscillator under phase diffusion, modeled here by a stochastic, zero-mean Gaussian-distributed phase $\theta$ \cite{Parisexp}. It is assumed that the initial state of the probe $\hat\rho=\vert\psi\rangle\langle\psi\vert$ is pure. Under these conditions, the state of the probe, after undergoing the $\phi$-dependent nonunitary process, becomes:
\begin{equation}
\hat\rho(\phi)=\int_{\mathbb{R}}g(\theta;\beta)e^{-i(\phi+\theta)\hat{a}^{\dagger}\hat{a}}\hat\rho e^{i(\phi+\theta)\hat{a}^{\dagger}\hat{a}}d\theta,
\end{equation}
where $g\!(\theta;\beta)\!=\!\exp{\![-\theta^{2}\!/\!(4\beta^2)]}/\sqrt{4\pi\beta^2}$, $\beta$ quantifies the degree of diffusion, and $\hat{a}$ ($\hat{a}^{\dagger}$) is the annihilation (creation) operator.

For this problem, an analytical expression of the noise-to-sensibility ratio is calculated by taking the initial state of the probe as a (real) Gaussian state, such that $\vert\psi\rangle=\hat{D}(\alpha)\hat{S}(r)\vert0\rangle$, where $\hat{S}\!(r)\!=\!\exp{\![r(\hat{a}^{\dagger2}\!-\!\hat{a}^{2})/2]}$ and $\hat{D}(\alpha)\!=\!\exp{\![\alpha(\hat{a}^{\dagger}\!-\!\hat{a})]}$ are the squeezing and displacement operators, respectively, with $\alpha$ and $r$ real numbers, and $\vert0\rangle$ the ground state of the oscillator, and by taking also $\hat{\cal M}(\phi_{\rm exp})\!=\!\hat{a}e^{i\phi_{\rm exp}}\!+\!\hat{a}^{\dagger}e^{-i\phi_{\rm exp}}$ as the observable to be measured, in which $\phi_{\rm exp}$ is an adjustable variable. From this analytical expression, one concludes that the best calibration, which decreases the noise-to-sensibility ratio as maximum as possible, happens for $\phi_{\rm exp}=\phi_{\rm true}-\pi/2$. This is equal to the one obtained without diffusion. Thus, this kind of noise does not change the optimum phase tune for these states. Furthermore, for this calibration, the effect of diffusion in the sensitivity is (i) to attenuate the sensibility by a factor of $\exp{\!(-\beta^2)}$, when compared with noiseless case, $2\alpha$; and (ii) to add the diffusion noise $[1\!-\exp{\!(-4\beta^2)}][2\alpha^2\!+\sinh{\!(2r)}]$ to the pure quantum one $\exp{\!(-2r)}$, which already exists in absence of diffusion. For these reasons, the information obtained about $\phi_{\rm true}$ with this strategy of estimation is
\begin{equation}\label{fbeta}
{\cal F}^{\rm nsr}(r,\alpha,\beta)\!=\!\dfrac{4\alpha^{2}e^{-2\beta^2}}{e^{-2r}+[1-e^{-4\beta^2}][2\alpha^2+\sinh{(2r)}]}.
\end{equation}

\begin{figure}[t]
\centering
\includegraphics[width=0.21\textwidth,height=0.20\textwidth]{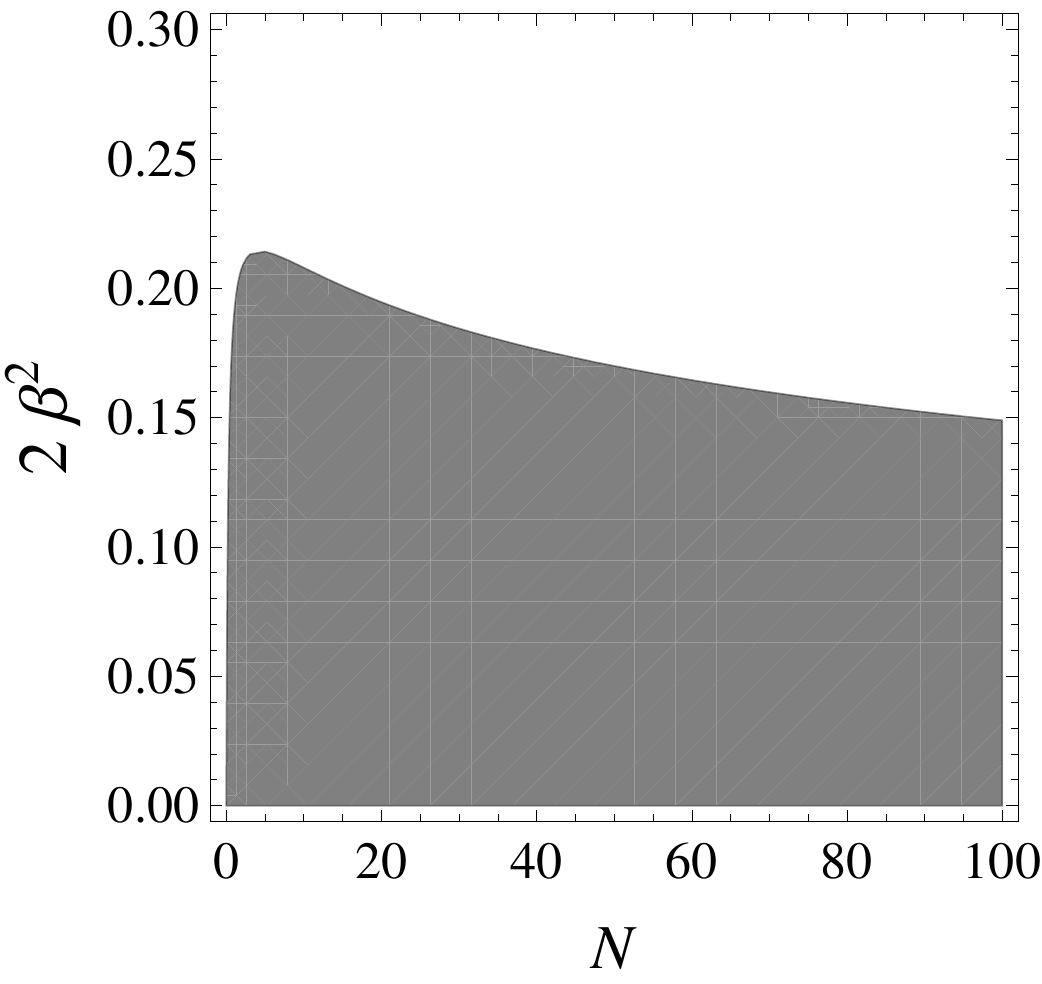}\hspace{0.5cm}\includegraphics[width=0.21\textwidth,height=0.21\textwidth]{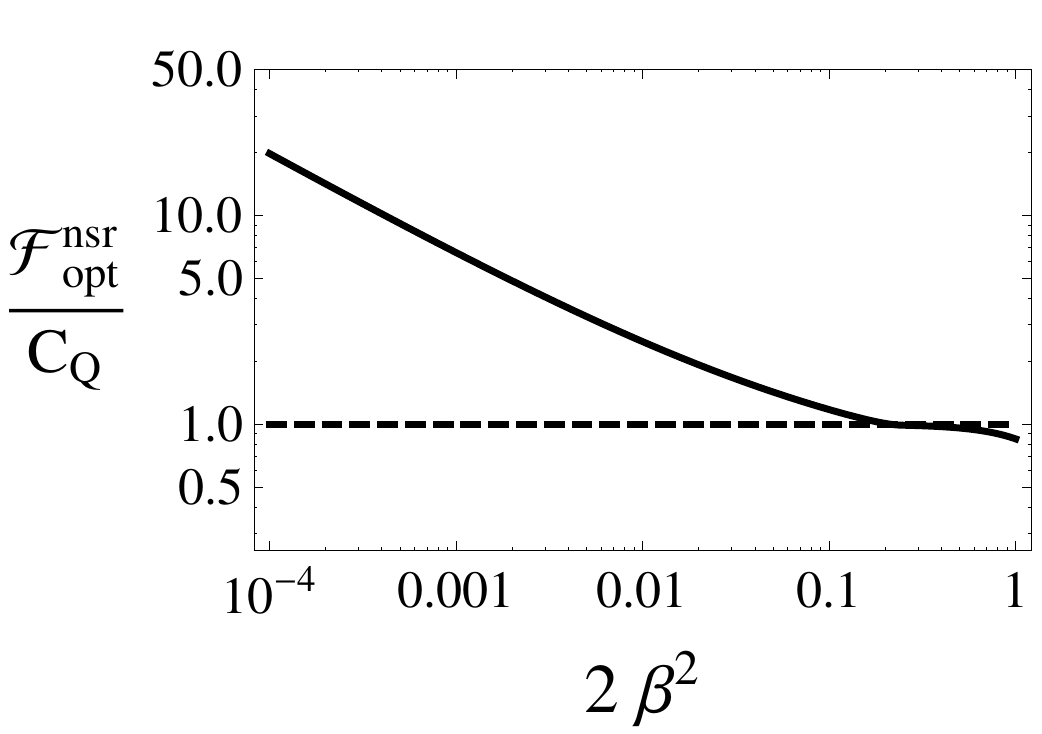}
\caption{Quantum sensitivity enhancement due squeezing. Left panel: shaded region on the plane ($N$, $2\beta^2$) marks the coordinates where there is some improvement in the sensitivity: ${\cal F}^{\rm nsr}_{\rm opt}/C_{\rm Q}\ge1$. Right panel: loglog plot of the maximum of ${\cal F}^{\rm nsr}_{\rm opt}/C_{\rm Q}$ over $N$ as a function of $2\beta^2$ (full line); dashed line is a benchmark for the standard sensitivity limit.}
\label{fig2}
\end{figure}

This quantity is therefore a lower bound to QFI for those Gaussian states. Being analytical, it gives also important information about the effect of diffusion on the sensitivity for all range of those parameters considered. For instance, Eq.~\eqref{fbeta} yields that it is possible to improve the sensitivity by squeezing the state of the probe, but only if $r\!\le\! r_{\rm max}$, where $\exp{\!(4r_{\rm max})}=\coth{\!(2\beta^2)}$. It may be understood physically by resorting to a pictorial representation of this state in the phase space: because of diffusion, if the ``length" of the state, which looks like a pointer of a clock, is larger than a threshold value, the probe will contribute actually to increase the noise, and then the sensitivity will decrease, since the squeezing does not improve the sensibility in this case. 

For a given average number $N=\alpha^2+\sinh^{2}\!(r)$ of quantum excitations, the optimum squeezing that maximizes ${\cal F}^{\rm nsr}$ in \eqref{fbeta} is given by $\exp{\!(2r_{\rm opt}\!)}=2{\cal N}\cosh{\!(2\beta^2)}/[1+\sqrt{1+2{\cal N}^{2}\sinh{\!(4\beta^2)}}]$, where ${\cal N}=(2N+1)\exp{(2\beta^2)}$. One can show then that the correspondent amount of information ${\cal F}^{\rm nsr}_{\rm opt}$ may surpass the value of ${\cal F}_{\rm std}$, obtained from the standard sensitivity limit under diffusion. The shaded region in the left panel of Fig.~\ref{fig2} shows the values of $2\beta^2$ and $N$ such that it happens for sure, since these results are obtained analytically by using the upper bound $C_{\rm Q}=4N/[1+8\beta^2N]$ to ${\cal F}_{\rm std}$ \cite{Escherxiv12}. While the right panel of Fig.~\ref{fig2} shows, as a function of $2\beta^2$, the maximum of the fraction ${\cal F}^{\rm nsr}_{\rm opt}/C_{\rm Q}$ over $N$, which already guarantees some quantum enhancement for the sensitivity for $2\beta^2\le0.21$ and for some finites $N$.  Notice that, for $\beta=0$, the noiseless case, ${\cal F}^{\rm nsr}_{\rm opt}=4N(N+1)$ is recovered, while for $2N[1-\exp{(-4\beta^2)}]\gg1$, ${\cal F}^{\rm nsr}_{\rm opt}\to{\rm csch}(2\beta^2)$, being, for $\beta\lesssim1$, just slight different to the asymptotic ($N\to\infty$) upper bound to QFI, $1/(2\beta^2)$~\cite{Escherxiv12}.

On the other hand, Eq.~\eqref{fbeta}, together with $C_{\rm Q}$, yields that the fraction of information obtained without squeezing over the standard one obeys ${\cal F}^{\rm nsr}/{\cal F}_{\rm std}\ge[1+8\beta^{2}N]/[\exp{(2\beta^2)}+4\sinh{(2\beta^2)}N]$, which is still close to the unity, even for moderate phase diffusion ($\beta\lesssim1$). For instance, taking $\beta=0.63$, this bound changes monotonically from $73\%$ to $90\%$ as $N$ increases. Therefore, based on these analytical results, one concludes that, with this strategy, the estimation of $\phi_{\rm true}$ is optimal for $\beta=0$ and it remains almost optimal until $\beta\sim1$. It increases the regime of utility shown in Ref.~\cite{Parisprl11} for homodyne detection to estimate a phase-shift under phase diffusion, which can, in fact, be tested experimentally \cite{Parisexp}.

\textit{Summary.}---I have presented in this Letter a prescription for evaluating the quantum sensitivity in single-parameter estimation. This approach was based on the sensitivities obtained by measuring physical observables, and, as a result of this analysis, it was shown that the symmetric logarithmic derivative operator, when well calibrated, yields to the minimal sensitivity allowed by quantum physics. The problem of how to achieve this quantum limit is reduced thus to a calibration issue. Thereupon, I have found the number of necessary trials to calibrate theoretically this optimal measurement device. These results were applied to the problem of phase-shift estimation under phase diffusion, where it was shown that squeezing and homodyne detection still allow quantum enhancements, surpassing the correspondent standard sensitivity limit, even for moderated phase diffusion. To conclude, I hope these results might be very useful from a theoretical point of view of quantum metrology, and might also inspire further developments in experiments aimed to overcome standard precision limits.
 
The author acknowledges financial support from the Brazilian funding agency CNPq.  This work was performed as part of the Brazilian National Institute for Science and Technology on Quantum Information.


\begin{thebibliography}{99}
\bibitem{com1} QFI may be given through the symmetric logarithmic derivative operator. However, to evaluate this operator, it is required to find the eigenvalue decomposition of the mixture probe state as a function of the parameter to be estimated \cite{Braunsprl94}.
\bibitem{estc} The considered estimator, for a given experimental observation $\langle\hat{\cal M}\rangle_{\rm exp}$, is defined formally as the value $x_{\rm est}$ such that the equality $\langle\hat{\cal M}\rangle_{x_{\rm est}}=\langle\hat{\cal M}\rangle_{\rm exp}$ is verified.
\bibitem{inec} Using a sample of $\nu$ elements, one can estimate $x_{\rm true}$ and use it to calibrate the optimal experimental device. Therefore, in average, $\delta x_{\rm exp}$ should also be limited by the quantum sensitivity.
\end{thebibliography}
\end{document}